\documentclass[aps,pra,twocolumn,groupedaddress]{revtex4-1}
\usepackage{amsmath}
\usepackage{xcolor}
\usepackage{commath}
\usepackage{graphicx}
\usepackage{hyperref}
\usepackage[capitalise]{cleveref}
\DeclareMathOperator{\tr}{tr}

\newcommand{\delnospace}[1]{\mathopen{}\del{#1}}
\begin{document}

\title{Sequential nonabsorbing microwave single-photon detector}

\author{Ivan Iakoupov}
\email[ivan.las@tmd.ac.jp; Current address: College of Liberal Arts and Sciences, Tokyo Medical and Dental University, 2-8-30 Kounodai, Ichikawa, Chiba 272-0827, Japan]{}
\author{Yuichiro Matsuzaki}
\email[Current address: Device Technology Research Institute, National Institute of Advanced Industrial Science and Technology (AIST), 1-1-1 Umezono, Tsukuba, Ibaraki 305-8568, Japan]{}
\author{William J. Munro}
\author{Shiro Saito}
\affiliation{NTT Basic Research Laboratories, NTT Corporation, 3-1 Morinosato-Wakamiya, Atsugi, Kanagawa 243-0198, Japan}

\date{\today}

\begin{abstract}
We propose a nonabsorbing microwave single-photon detector that uses an 
artificial atom as a coherent interaction mediator between a traveling photon 
and a high-$Q$ resonator, fully exploiting the knowledge of the photon's 
arrival time. Our proposal can be implemented with the current level of 
technology and achieves distinguishability (probability of distinguishing 
between zero and one photon) in excess of 98\% for realistic parameters. This 
is better than any of the similar detector proposals, even the ones using 
several artificial atoms.
\end{abstract}


\maketitle
\section{Introduction}
Efficiently detecting traveling microwave photons is an extremely challenging 
yet important task for future quantum technology. Currently, it is still not 
clear what is the best approach or whether there even \emph{is} an approach 
that can be used in all situations. Out of the multitude of different 
proposals, we will restrict our attention here to the more versatile 
nonabsorbing detectors, so that detectors that absorb without reemitting or 
reemit the photon at a different
frequency~\cite{romero_prl09,chen_prl11,koshino_prl13,koshino_pra15,inomata_ncomms16,kyriienko_prl16,oelsner_prapl17,leppakangas_pra18,lescanne_prx20} 
are out of scope of this article. There is a number of recent proposals for 
the nonabsorbing microwave single-photon 
detectors~\cite{helmer_pra09,fan_prl13,sathoyamoorthy_prl14,fan_prb14,royer_prl18} 
that can, in principle, be operated continuously, in the sense that the 
detector could be continuously interrogated for the presence of a photon. This 
way, both the presence of a photon and its arrival time could be obtained. 
The other type of the nonabsorbing detectors is the one that relies on 
knowledge of the photon's arrival time (or an arrival time window), and this 
is the approach taken in the recent experimental 
realizations~\cite{kono_nphys18,besse_prx18}. We will call this type of the 
detectors sequential, since their principle of operation usually consists of a 
sequence of different operations, where the simplest version involves two 
distinct steps: 1.~interaction between the incident photon and the detector, 
2.~interrogation of the detector. For instance, in 
Refs.~\cite{kono_nphys18,besse_prx18}, step 1 is the controlled-phase gate 
between the photon and an artificial atom~\cite{duan_prl04}. Often, trying to 
perform step~2 simultaneously with step~1 results in the detector not working 
at all, since interrogation will prevent the interaction. The continuous-mode 
detectors have various approaches to make the interrogation during the 
interaction time be less 
detrimental~\cite{sathoyamoorthy_prl14,fan_prb14,royer_prl18}.

The line between the continuous-mode and the sequential detectors is often 
blurred, however, as the continuous-mode detectors can also be operated 
sequentially where the sequence consists of ``interrogation off'' and 
``interrogation on''. This is the mode that is used for the calculation of one 
of the often used figures of merit for the single-photon detectors -- the 
distinguishability (also known as ``measurement fidelity''), which is the 
probability of correctly distinguishing between zero and one photon. 
Operationally, this is because the contribution of the signal due to an 
incident photon is finite, but the contribution of the (quantum and technical) 
noise can always be increased by making the interrogation time longer. 
Therefore, the best distinguishability value is obtained when the 
interrogation starts from the arrival time of the incident photon and 
continues until some later time that is determined by the response of the 
detector. In some cases, the proposals, which we call continuous-mode here, 
are not even analysed as such, and only their sequential mode of operation is 
quantified by calculating the distinguishability~\cite{fan_prb14, 
sathoyamoorthy_prl14}. Even for the cases when the quantitative 
continuous-mode analysis is done~\cite{royer_prl18}, the distinguishability is 
usually playing a very prominent role in the analysis. The reported 
distinguishabilities are in the range 70\%-96\%. Compared to those proposals, 
our proposed detector can no longer be operated in a continuous mode but 
achieves higher distinguishability with less experimental complexity (in terms 
of the number of artificial atoms).

In the telecom wavelengths, operation of the continuous-mode detectors in the 
sequential mode (usually called the gated mode) is widely practiced in quantum 
communication~\cite{gisin_rmp02,zhang_light15} to reduce 
noise~\cite{ribordy_appl_opt_98}. In general, the photon's arrival time at the 
detector can be determined whenever the photon's emission time is known, and 
the path length between the source and the detector is constant. Both 
assumptions hold true for a typical setup involving microwave photons and 
superconducting circuits, where deterministic single-photon sources are 
available~\cite{forn_diaz_prapplied17,zhou_prapplied20}, and the photons are 
routed by fixed transmission lines. Hence, sequential operation is usually 
possible, and our results suggest that a sequential detector that is designed 
to fully exploit the knowledge of the photon's arrival time is preferable over 
the continuous-mode detectors that are operated in the sequential mode.

\begin{figure}[t]
\begin{center}
\includegraphics{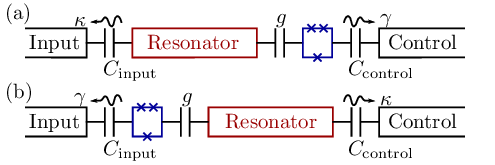}

\includegraphics{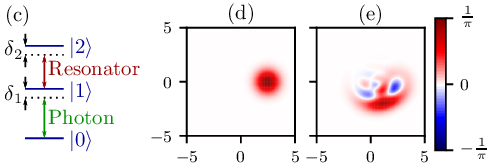}

\includegraphics{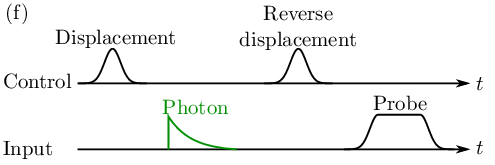}
\end{center}
\caption{(a) The usual microwave circuit QED setup. Here, with a 
persistent-current artificial atom~\cite{orlando_prb99} and capacitive 
couplings, but many other variations are possible. The coupling capacitance 
$C_\text{control}$ is relatively small to prevent the artificial atom from 
decaying into the control transmission line (right). The coupling capacitance 
$C_\text{input}$ is relatively large to facilitate the interaction of the 
fields in the input transmission line (left) with the resonator. The decay 
rates are indicated next to the respective capacitive coupling that gives rise 
to them ($\gamma$ means both decay rates $\gamma_{01}$ and $\gamma_{12}$). 
(b) The setup employed for our proposed detector. The coupling capacitances 
are the same, but the places and the roles of the qubit and the resonator are 
reversed. (c) The level diagram of the artificial atom and indication of the 
system that the given transition is coupled to (either the incident photon or 
the resonator). (d) Wigner function of the initial coherent state of the 
resonator. The average photon number is 
$\langle \hat{a}^\dagger\hat{a}\rangle = 3$~\cite{fan_prb14}. (e) Wigner 
function of the resonator state (reduced density matrix) after the interaction 
with an incident photon with bandwidth $\gamma_c/\gamma_{01}=0.1$ for a time 
$T_\text{interact}\approx 92/\gamma_{01}$. The other parameters are: 
$\kappa/\gamma_{01}=3.2\cdot 10^{-5}$, 
$\gamma_{11}/\gamma_{01}=3.2\cdot 10^{-3}$, 
$\gamma_{22}/\gamma_{01}=6.4\cdot 10^{-3}$, $\gamma_{12}/\gamma_{01}=0.1$, 
$\delta_1/\gamma_{01}\approx -1.380$, $\delta_2/\gamma_{01}\approx -96.89$, 
$g/\gamma_{01}\approx 7$. $T_\text{interact}$, $\delta_1$ and $\delta_2$ were 
chosen to minimize the error probability $P_{\text{E},M}$, giving 
$P_{\text{E},M}\approx P_{\text{E,opt}}\approx 2.2\%$. (f) The proposed 
detection sequence for the setup in (b). The single-photon field is green and 
classical drives (displacements and probe) are black. They are incident from 
the two different transmission lines (``Control'' and ``Input'').
\label{fig_setup}}
\end{figure}

We believe that our proposed detector could also be competitive against the 
existing experimental realizations of the sequential 
detectors~\cite{kono_nphys18,besse_prx18} where an artificial atom is used to 
store a state that is measured after an initial interaction with the incident 
photon, mediated by the resonator (see Fig.~\ref{fig_setup}(a)). The proposed 
detector uses the resonator to store a state that is measured after an initial 
interaction with the incident photon, mediated by the artificial atom (see 
Fig.~\ref{fig_setup}(b)). Thereby, our proposed detector reverses the roles of 
the artificial atom and the resonator. Technical details complicate the answer 
with respect to which approach is better (both give distinguishability 
$\mathcal{F}=100\%$ in the idealized limit), but it is usually easier to 
achieve long coherence time in 
resonators~\cite{reagor_prb16,wang_science16,romanenko_prapplied20} than in 
artificial atoms~\cite{burnett_npjqi19}. Therefore, our proposed detector may 
be less susceptible to the errors due to the finite coherence time of the 
stored state, which is one of the main sources of 
error~\cite{kono_nphys18,besse_prx18}.

\section{Setup}

The setup of our proposed detector is shown in Fig.~\ref{fig_setup}(b). It 
consists of a three-level artificial atom and a resonator. The artificial atom 
is coupled both to an input transmission line and the resonator. From the 
input transmission line, a single photon is incident with a carrier frequency 
that is detuned from the $|0\rangle\leftrightarrow|1\rangle$ transition 
by~$\delta_1$ (see Fig.~\ref{fig_setup}(c)). The resonator is close in 
frequency to the transition $|1\rangle\leftrightarrow |2\rangle$, resulting in 
a coupling $g$ with detuning $\delta_2$. The resonator frequency is assumed to 
be far detuned from the transition $|0\rangle\leftrightarrow|1\rangle$ such 
that the coupling is negligible. The resonator is also assumed to not directly 
couple to the input transmission line, e.g., due to suppression of the 
coupling by a Purcell filter~\cite{reed_apl10} at the resonator frequency 
attached to the input transmission line (left in Fig.~\ref{fig_setup}(b), not 
shown). Depending on the bandwidth, the same or a different Purcell filter 
could also suppress the decay from the state $|2\rangle$ of the artificial 
atom, resulting in a small decay rate $\gamma_{12}/\gamma_{01}=0.1$ that was 
assumed in Ref.~\cite{fan_prb14}. We keep the same $\gamma_{12}/\gamma_{01}$ 
for an easier comparison. The resonator is also coupled to a control 
transmission line with decay rate $\kappa$, making it possible to perform 
displacements of the resonator field.

The setup described above is very similar to the setup of 
Ref.~\cite{fan_prb14}, but with important differences, as described below. The 
Hamiltonian for our setup in the interaction picture with the rotating wave 
approximation made is
\begin{gather}
\label{Hamiltonian}
H=\hbar\delta_1\hat{\sigma}_{11}+\hbar(\delta_1+\delta_2)\hat{\sigma}_{22}
-i\hbar g(\hat{a}\hat{\sigma}_{21}-\hat{a}^\dagger\hat{\sigma}_{12}),
\end{gather}
where $\hat{\sigma}_{\mu\nu}=|\mu\rangle\langle\nu|$ are the atomic operators, 
and $\hat{a}$ is the annihilation operator of the resonator. Compared to the 
Hamiltonian of Ref.~\cite{fan_prb14}, the above Hamiltonian does not have the 
term $-i\hbar E(\hat{a}-\hat{a}^\dagger)$ that describes the always-on drive 
of the resonator. This is because in our proposal, the Rabi frequency $E$ is 
set to zero during the entire detection sequence, except for the displacements 
that assume a large enough $E$ for the displacements to happen in a time much 
shorter than any other time scale.

The incident photon that arrives from the input transmission line is modeled 
by a source resonator with decay rate
$\gamma_c$~\cite{fan_prl13,sathoyamoorthy_prl14,fan_prb14,royer_prl18}, 
setting the mode shape of the photon to be a decaying exponential in time. The 
corresponding master equation is~\cite{fan_prb14}
\begin{gather}
\label{deterministic_master_equation}
\begin{aligned}
\dot{\rho} 
= \mathcal{L}\rho
= &-\frac{i}{\hbar}[H,\rho] 
+\gamma_c\mathcal{D}[\hat{c}]\rho
+\gamma_{01}\mathcal{D}[\hat{\sigma}_{01}]\rho\\
&+\gamma_{12}\mathcal{D}[\hat{\sigma}_{12}]\rho
+\kappa\mathcal{D}[\hat{a}]\rho\\
&+\sqrt{\gamma_c\gamma_{01}}([\hat{c}\rho,\hat{\sigma}_{10}]
+[\hat{\sigma}_{01},\rho\hat{c}^\dagger]),
\end{aligned}
\end{gather}
where 
$\mathcal{D}[\hat{r}]\rho=\frac{1}{2}(2\hat{r}\rho\hat{r}^\dagger-\rho\hat{r}^\dagger\hat{r}-\hat{r}^\dagger\hat{r}\rho)$,
$\hat{c}$ is the annihilation operator of the source resonator, and the terms 
proportional to $\sqrt{\gamma_c\gamma_{01}}$ describe the coupling of the 
source resonator to the artificial atom using the cascaded systems 
formalism~\cite{gardiner_prl93,carmichael_prl93}. Setting $\kappa$ to be much 
smaller than in Ref.~\cite{fan_prb14} (which used $\kappa/\gamma_{01}=0.037$) 
is the key difference of our proposal. Ideally, $\kappa\rightarrow 0$ can be 
taken, corresponding to the limit of $\kappa$ being negligible on the time 
scales of the detection sequence. Such a choice of $\kappa$ prevents the 
continuous-mode operation (possible in the proposal of Ref.~\cite{fan_prb14}), 
because the resonator state cannot be probed directly. Instead, we propose a 
detection sequence (Fig.~\ref{fig_setup}(f)) whose final step is to probe the 
artificial atom to gain the information about the resonator state. Before we 
go in the details about this procedure, we can explain the high 
distinguishability of the detector by considering the distinguishability of 
the resonator states for the cases with and without an incident photon.

The detection sequence starts by initializing the artificial atom in state 
$|0\rangle$ and the resonator in a coherent state. The latter could be 
accomplished by initializing in a vacuum state and displacing it. We choose 
the average photon number $\langle \hat{a}^\dagger\hat{a}\rangle = 3$ 
\cite{fan_prb14}, resulting in the coherent state of Fig.~\ref{fig_setup}(d). 
At this point, an incident photon will change the state into the one shown in 
Fig.~\ref{fig_setup}(e) after an interaction time $T_\text{interact}$. We 
believe that the interaction is related to the selective number-dependent 
arbitrary phase (SNAP) gates~\cite{heeres_prl15}, but with the difference that 
our scheme uses a single photon rather than a classical field to drive the 
interaction. If no photon is incident and $\kappa\rightarrow 0$, the resonator 
state stays the same, because the coupling of the transition 
$|0\rangle\leftrightarrow|1\rangle$ of the artificial atom and the resonator 
is assumed to be negligible. For $\kappa > 0$, the only change is a decay of 
the resonator photon number, i.e., a displacement towards the vacuum state. 
Since a complete displacement towards the vacuum state is part of the 
detection protocol (as explained below), the error due to this decay can be 
compensated for.

We use two different distinguishability measures for the reduced density 
matrices of the resonator, $\rho_0$ and $\rho_1$, that correspond to the vacuum 
and single-photon inputs, respectively. The first one is the 
distinguishability $\mathcal{F}_\text{opt}$ that uses an (unspecified) optimal 
measurement and assumes a $50/50$ probability of the two 
states. It can be written $\mathcal{F}_\text{opt}=1-P_\text{E,opt}$, 
where~\cite{fuchs_ieee99}
\begin{gather}
P_\text{E,opt}=\frac{1}{2}-\frac{1}{4}\enVert{\rho_0-\rho_1}_\text{tr},
\end{gather}
with $\enVert{\rho_0-\rho_1}_\text{tr}$ being the trace norm of $\rho_0-\rho_1$.
For the second distinguishability measure, we note that we can write 
$\rho_0=|\alpha\rangle\langle\alpha|$ for some coherent state $|\alpha\rangle$, 
which is either the initial state shown in Fig.~\ref{fig_setup}(d) 
(for $\kappa\rightarrow 0$) or a displaced version of it (for $\kappa > 0$). We 
define two measurement operators, 
$M_0=|\alpha\rangle\langle\alpha|$ and $M_1=I-M_0$, where $I$ is the identity 
operator. Then the distinguishability is $\mathcal{F}_M=1-P_{\text{E},M}$, 
where~\cite{geremia_pra04}
\begin{gather}
P_{\text{E},M}
=\frac{1}{2}\tr\delnospace{\rho_0 M_1}
+\frac{1}{2}\tr\delnospace{\rho_1 M_0},
\end{gather}
again assuming a $50/50$ probability of the two states $\rho_0$ and $\rho_1$. 
This second state distinguishability measure is an idealization of the more 
realistic measurement procedure that we will discuss later. Since $M_0=\rho_0$, 
the term $\tr\delnospace{\rho_0 M_1}$ that represents the dark 
count probability in $P_{\text{E},M}$ vanishes, resulting in 
$P_{\text{E},M}=\tr\delnospace{\rho_1 M_0}/2$.

To model the pure dephasing of the artificial atom, we add the terms 
$\gamma_{11}\mathcal{D}[\hat{\sigma}_{11}]\rho$ and 
$\gamma_{22}\mathcal{D}[\hat{\sigma}_{22}]\rho$ to the master 
equation~\eqref{deterministic_master_equation} and set 
$\gamma_{22}/\gamma_{11}=2$ to account for the larger dephasing rate of the 
higher energy levels~\cite{peterer_prl15}. Assuming the decay rate 
$\gamma_{01}=2\pi\cdot10\text{ MHz}$ and the pure dephasing time 
$T_{\phi,\text{AA}}=10\text{ }\mu\text{s}$ for the artificial atom, we have 
$\gamma_{11}/\gamma_{01}=2/(T_{\phi,\text{AA}}\gamma_{01})\approx 3.2\cdot 10^{-3}$
 and $\gamma_{22}/\gamma_{01}\approx 6.4\cdot 10^{-3}$. For the resonator, we 
assume $T_{1,\text{R}}=500\text{ }\mu\text{s}$~\cite{axline_nphys18}, giving 
$\kappa/\gamma_{01}=1/(T_{1,\text{R}}\gamma_{01})\approx 3.2\cdot 10^{-5}$. 
Even longer coherence times are possible with the 3D 
resonators~\cite{reagor_prb16,wang_science16,romanenko_prapplied20}. For 
reference, we also compare to $T_{1,\text{R}}=50\text{ }\mu\text{s}$ that 
represents the 2D resonators~\cite{megrant_apl12}. The other imperfections in 
the model are the non-zero ratio $\gamma_c/\gamma_{01}$ and the imperfect 
dispersive interaction. The latter depends on how close the 
Hamiltonian~\eqref{Hamiltonian} is to the perfect dispersive Hamiltonian
\begin{gather}
\label{Hamiltonian_disp}
H_\text{disp}=\hbar\delta_1\hat{\sigma}_{11}
-\hbar\chi\hat{\sigma}_{11}\hat{a}^\dagger\hat{a},
\end{gather}
where $\chi = g^2/(\delta_1+\delta_2)$. We choose 
$g=7\gamma_{01}=2\pi\cdot70\text{ MHz}$~\cite{axline_nphys18}.

Lossy elements like circulators in the path between the source and the detector can be 
described by fictitious beam splitters~\cite{sathoyamoorthy_prl14}. A beam 
splitter with the transmission coefficient $t$ corresponds to the power loss 
$1-t^2$. By using the beam splitter relations and tracing over the loss mode, 
the final state $\rho_1$ (single-photon input) is modified to 
$t^2\rho_1+(1-t^2)\rho_0$, while the final state $\rho_0$ (vacuum input) is 
left unchanged. The error probability $P_{\text{E},M}$ with a lossy element 
can be calculated in a particularly simple way due to
$M_0=\rho_0$ (perfect compensation for $\kappa > 0$). We get 
$P_{\text{E},M}=(1-t^2)/2+t^2\tr\delnospace{\rho_1 M_0}/2$. For small power 
loss, we can approximately say that half of the power loss $1-t^2$ gets added 
to the intrinsic error probability of the detector.

\section{Results}

\begin{figure}[t]
\begin{center}
\includegraphics{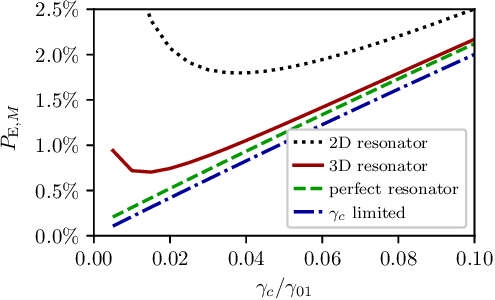}
\end{center}
\caption{Error probabilities $P_{\text{E},M}$ as a function of the input 
photon bandwidth $\gamma_c$. Common parameters: $\gamma_{12}/\gamma_{01}=0.1$, 
$g/\gamma_{01}=7$. Unless noted otherwise, Hamiltonian~\eqref{Hamiltonian} is 
used, and we set $\gamma_{11}/\gamma_{01}=3.2\cdot 10^{-3}$ and 
$\gamma_{22}/\gamma_{01}=6.4\cdot 10^{-3}$. The legend indicates the different 
scenarios: ``2D resonator'' uses
$\kappa=3.2\cdot 10^{-4}\gamma_{01}$ ($T_{1,\text{R}}=50\text{ }\mu\text{s}$), 
``3D resonator'' uses $\kappa=3.2\cdot 10^{-5}\gamma_{01}$ 
($T_{1,\text{R}}=500\text{ }\mu\text{s}$), ``perfect resonator'' uses 
$\kappa=0$, and ``$\gamma_\text{c}$~limited'' uses
Hamiltonian~\eqref{Hamiltonian_disp} with $\kappa=\gamma_{11}=\gamma_{22}=0$. For each 
value of $\gamma_c/\gamma_{01}$, the parameters $\delta_1$, $\delta_2$ and 
$T_\text{interact}$ were chosen to minimize $P_{\text{E},M}$.
}
\label{fig_gammaC}
\end{figure}

In Fig.~\ref{fig_gammaC}, we vary $\gamma_c$ and plot $P_{\text{E},M}$ for 
optimal $\delta_1$ and $\delta_2$ using both Hamiltonians $H$ and 
$H_\text{disp}$. We see that the optimal detunings are such that 
$\chi=g^2/(\delta_1+\delta_2)\approx -0.5\gamma_{01}$ with a small change for 
the considered range of $\gamma_c$. This shows that the ratio between the 
decay rate of the artificial atom and the dispersive shift should be 
(approximately) fixed, similar to Ref.~\cite{kono_nphys18}, where the decay 
rate of the resonator and the dispersive shift needed to have a fixed ratio. 
Additionally, $\delta_1/\chi\approx2.75$ for the considered range of $\gamma_c$. 
It is close to $\langle\hat{a}^\dagger\hat{a}\rangle=3$, which is the average 
number of photons in the resonator state. This shows that the primary function 
of the detuning $\delta_1$ is to compensate for the static shift of the 
frequency of the artificial atom due to the dispersive interaction 
$-\chi\hat{\sigma}_{11}\langle\hat{a}^\dagger\hat{a}\rangle$. The interaction 
time $T_\text{interact}$ is varied such that 
$T_\text{interact}\leq 10/\gamma_c$ with the smallest $P_{\text{E},M}$ is 
chosen. E.g., for $\kappa/\gamma_{01}=3.2\cdot 10^{-5}$ and 
$\gamma_c/\gamma_{01}=0.1$, 
$T_\text{interact}\approx 92/\gamma_{01}\approx 1.5\text{ }\mu\text{s}$ is 
chosen. For the optimized parameters, $P_{\text{E,opt}}\approx P_{\text{E},M}$, 
which would have lead to completely overlapping curves in 
Fig.~\ref{fig_gammaC} (very small difference is visible for 
$\kappa/\gamma_{01}=3.2\cdot 10^{-4}$). This is not always the case: if we 
used exactly the same parameters as in Ref.~\cite{fan_prb14} except setting 
$\kappa=0$ ($\gamma_c/\gamma_{01}=0.1$, $\gamma_{11}=\gamma_{22}=0$, 
$\delta_1/\gamma_{01}=-0.8$, $\delta_2/\gamma_{01}=-18$, $g/\gamma_{01}=2.45$) 
then $P_{\text{E},M}\approx 6.7\%$ and $P_\text{E,opt}\approx 4.6\%$.

The solid red curve in Fig.~\ref{fig_gammaC} shows our results with the most 
complete model of the imperfections that uses the 
Hamiltonian~\eqref{Hamiltonian} with $\kappa/\gamma_{01}=3.2\cdot 10^{-5}$. 
The states in Figs.~\ref{fig_setup}(d)~and~(e) correspond to 
$\gamma_c/\gamma_{01}=0.1$ on this curve. The dotted black and dashed green 
curves only differ from the solid red one in setting 
$\kappa/\gamma_{01}=3.2\cdot 10^{-4}$ and $\kappa=0$, respectively. The 
dash-dotted blue curve represents the almost ideal case with 
$\kappa=\gamma_{11}=\gamma_{22}=0$ and uses the 
Hamiltonian~\eqref{Hamiltonian_disp} so that it is only limited by the 
non-zero ratio $\gamma_c/\gamma_{01}$, achieving $\mathcal{F}_M=100\%$ for 
$\gamma_c/\gamma_{01}\rightarrow 0$. Comparison of these curves illustrates 
the different effect of the imperfections of the artificial atom and the 
resonator. The imperfections of the artificial atom add a constant term to 
$P_{\text{E},M}$, since the artificial atom only acts as a mediator of the 
interaction and effectively stores every part of the incident photon only for 
a limited time. The non-zero $\kappa$ of the resonator, on the other hand, 
gives a larger error with smaller $\gamma_c/\gamma_{01}$, since the state has 
to be stored in the resonator for a longer time.

A realization of the measurement used for $\mathcal{F}_M$ starts with an 
unconditional displacement of $\rho_0$ and $\rho_1$ such that
$\rho_0=|\text{vac}\rangle\langle\text{vac}|$, where $|\text{vac}\rangle$ is the vacuum 
state of the resonator, and the ideal measurement operators become 
$M_0=|\text{vac}\rangle\langle\text{vac}|$ and $M_1=I-M_0$. An implementation 
of these measurement operators can be done by driving the artificial atom by a 
continuous wave field with a frequency close to the transition 
$|0\rangle\leftrightarrow|1\rangle$ from the input transmission line (using a 
switch~\cite{pechal_prl16,besse_prx18} to change from the single-photon 
source) and doing homodyne detection of the reflected field. The idea is that 
the dispersive interaction of the artificial atom with the resonator (see 
Eq.~\eqref{Hamiltonian_disp}) shifts the frequency of the transition 
$|0\rangle\leftrightarrow|1\rangle$ only if the resonator is in a state 
different from vacuum. This conditional frequency shift of the transition 
$|0\rangle\leftrightarrow|1\rangle$ gives a conditional phase shift of the 
reflected probe field applied at a fixed frequency. The complete detection 
sequence is summarized in Fig.~\ref{fig_setup}(f).

The probing of the artificial atom is described by an additional Hamiltonian 
term $\hbar\Omega(\hat{\sigma}_{01}+\hat{\sigma}_{10})$, where $\Omega$ is the 
Rabi frequency of the continuous wave drive, and using the stochastic master 
equation~\cite{fan_prl13,sathoyamoorthy_prl14,fan_prb14,royer_prl18}
\begin{gather}
d\rho=\mathcal{L}\rho dt + \sqrt{\eta}dW(t)\mathcal{H}[\hat{O}]\rho,
\end{gather}
with $\mathcal{H}[\hat{r}]\rho=\hat{r}\rho+\rho\hat{r}^\dagger
-\tr\delnospace{\hat{r}\rho+\rho\hat{r}^\dagger}\rho$,
$\hat{O}=e^{-i\phi}\sqrt{\gamma_{01}}\hat{\sigma}_{01}$, $\eta$ being the 
efficiency of the homodyne detection ($\eta=1$ unless noted otherwise), and
$dW(t)$ being a Wiener process. Since the interaction time is long relative to 
the duration of the incident photon, the atom is negligibly different from 
being in state $|0\rangle$ at the beginning of probing. The initial system 
state for probing is thus taken to be a product state of the artificial atom 
being in state $|0\rangle$, and the resonator being in the state determined by 
the interaction (with and without an incident photon) and the reverse 
displacement.

The corresponding homodyne current is
\begin{gather}
I_n(t)=\sqrt{\eta}\langle \hat{O}+\hat{O}^\dagger\rangle + dW(t)/dt,
\end{gather}
where the subscript $n$ can be 1 or 0 -- indicating whether a photon was 
incident or not, respectively. To filter this homodyne current, we define 
$\bar{I}_n(t)=\langle \hat{O}+\hat{O}^\dagger\rangle$ where the expectation 
value uses the density matrix evolved with the deterministic master 
equation~\eqref{deterministic_master_equation}. The filtered integrated 
homodyne current is then
\begin{gather}
S_n=\int_{T_\text{interact}}^{T_\text{interact}+T_\text{probe}} I_n(t)h(t)dt,
\end{gather}
where the filter is $h(t)=|\bar{I}_0(t)-\bar{I}_1(t)|$, and $T_\text{probe}$ 
is the probing time. We generate $N_\text{tot}=10^4$ 
trajectories for each of the cases $n=0$ and $n=1$, resulting in the integrals 
$S_{n,j}$, where $j$ is the trajectory index.

\begin{figure}[t]
\begin{center}
\includegraphics{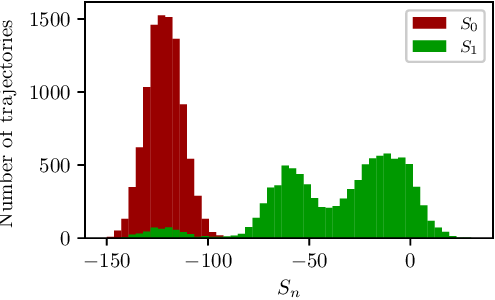}
\end{center}
\caption{Histogram of the filtered integrated homodyne currents, $S_0$ and 
$S_1$, that correspond to the case without and with an incident photon, 
respectively. The parameters correspond to $\gamma_c/\gamma_{01}=0.1$ on the 
solid red curve in Fig.~\ref{fig_gammaC} ($\kappa/\gamma_{01}=3.2\cdot 10^{-5}$, 
$\gamma_{11}/\gamma_{01}=3.2\cdot 10^{-3}$, 
$\gamma_{22}/\gamma_{01}=6.4\cdot 10^{-3}$, $\gamma_{12}/\gamma_{01}=0.1$, 
$g/\gamma_{01}=7$, $\delta_1/\gamma_{01}\approx -1.380$, 
$\delta_2/\gamma_{01}\approx -96.89$, $T_\text{interact}=92/\gamma_{01}$). The 
parameters for probing of the artificial atom are: $\delta_1=0.1\gamma_{01}$ 
$\phi=\pi/2$, $\Omega/\gamma_{01}=0.2$, $T_\text{probe}=500/\gamma_{01}$. The 
total number of the trajectories is $N_\text{tot}=10^4$. The resulting error 
probability is $P_{\text{E},M,\text{real}}\approx 2.4\%$.
}
\label{fig_sme_histogram}
\end{figure}

The distinguishability is then 
$\mathcal{F}_{M,\text{real}}=1-P_{\text{E},M,\text{real}}$, where (for a 
$50/50$ probability of the two states)
\begin{gather}
P_{\text{E},M,\text{real}}
=\frac{1}{2}\frac{N_{S_{0,j}>S_\text{thr}}}{N_\text{tot}}
+\frac{1}{2}\frac{N_{S_{1,j}<S_\text{thr}}}{N_\text{tot}},
\end{gather}
with $N_{S_{0,j}>S_\text{thr}}$ ($N_{S_{1,j}<S_\text{thr}}$) being the number 
of the integrals $S_{0,j}$ ($S_{1,j}$) that are above (below) the threshold 
$S_\text{thr}$. The threshold $S_\text{thr}$ is chosen such that 
$P_{\text{E},M,\text{real}}$ is minimized. This definition assumes that, on 
average, the integrals $S_{0,j}$ are smaller than the integrals $S_{1,j}$. It 
is the case in Fig.~\ref{fig_sme_histogram} which corresponds to 
$\gamma_c/\gamma_{01}=0.1$ on the solid red curve in Fig.~\ref{fig_gammaC}. 
The term $N_{S_{0,j}>S_\text{thr}}/N_\text{tot}$ in the definition of 
$P_{\text{E},M,\text{real}}$ is the dark count probability. We 
find that for the threshold $S_\text{thr}\approx -93$ that minimizes 
$P_{\text{E},M,\text{real}}$ in Fig.~\ref{fig_sme_histogram}, the term 
$N_{S_{0,j}>S_\text{thr}}/N_\text{tot}$ is negligible compared to 
$N_{S_{1,j}<S_\text{thr}}/N_\text{tot}$, and the resulting error 
probability $P_{\text{E},M,\text{real}}\approx 2.4\%$ is determined by the 
latter term. This implies that including the effect of lossy elements with 
power loss $1-t^2$ in the path between the source and the detector can be done 
in the same way as for $P_{\text{E},M}$ with the resulting expression 
$P_{\text{E},M,\text{real}}\approx(1-t^2)/2+t^2 N_{S_{1,j}<S_\text{thr}}/(2N_\text{tot})$.

The error probability $P_{\text{E},M,\text{real}}\approx 2.4\%$ is slightly 
larger than $P_{\text{E},M}\approx 2.2\%$ for the same interaction parameters. 
It might be possible to optimize the probing parameters (see caption of 
Fig.~\ref{fig_sme_histogram}) further to make the difference smaller. One of 
these parameters is the probing time $T_\text{probe}=500/\gamma_{01}$, which 
could be either increased to reduce the error probability or decreased to 
shorten the total detection time 
$T_\text{interact}+T_\text{probe}=592/\gamma_{01}=9.4\text{ }\mu\text{s}$. 
Imperfect homodyne detection efficiency $\eta=0.5$ increases the error 
probability to $P_{\text{E},M,\text{real}}\approx 3.1\%$, showing that our 
proposal is robust against a non-unit $\eta$.

\section{Discussion and Conclusion}

The distinguishability of our sequential detector proposal compares favorably 
against the distinguishability reported for the continuous-mode detector 
proposals~\cite{sathoyamoorthy_prl14,fan_prb14,royer_prl18}, even if they use 
several artificial atoms. Before the comparison, we note that all the 
considered setups are such that the photon to be 
detected is reflected back into the input transmission line. Therefore, a 
circulator is needed to prevent the reflected photon from going back to the 
source. Our proposal has the same behavior. In an experiment, the insertion 
loss of the circulator (around 4\% power loss~\cite{lnf_circulator}) will 
therefore affect the distinguishability, and we have explained above how such 
loss could be included in the theoretical analysis for our proposal. However, 
for an equivalent comparison with other 
proposals~\cite{sathoyamoorthy_prl14,fan_prb14,royer_prl18}, we assume perfect 
circulators, since the distinguishability numbers with lossy circulators are 
not provided in all references.

As shown in Fig.~\ref{fig_gammaC}, we report 
distinguishabilities $\mathcal{F}=98\%$ and above for 
$\gamma_c/\gamma_{01}\leq 0.1$, which are better than the reported 
distinguishabilities in 
Refs.~\cite{sathoyamoorthy_prl14,fan_prb14,royer_prl18} (70\%-96\%). In the 
other proposals, the artificial atom decay rates may have a range of 
values~\cite{sathoyamoorthy_prl14}, or the collective effects may make the 
single artificial atom rate less relevant than the decay rate of the entire 
ensemble~\cite{royer_prl18}. In the former case~\cite{sathoyamoorthy_prl14}, 
$\mathcal{F}=90\%$ is reported for 8~artificial atoms where 
$\gamma_c/\gamma_{01}$ varies between $0.5$ and $0.1$ for the different 
artificial atoms. In the latter case~\cite{royer_prl18}, 
$\gamma_c/\gamma_\text{B}=0.1$ is used, where $\gamma_\text{B}$ is the decay 
rate of the bright state of the ensemble, resulting in $\mathcal{F}=96\%$ for 
4~artificial atoms and 1~resonator. The proposal of Ref.~\cite{fan_prb14} uses 
$\gamma_c/\gamma_{01}=0.1$ and reports $\mathcal{F}=90\%$ for 2~artificial 
atoms and 2~resonators. Thus, our proposed detector achieves higher 
distinguishability with less experimental complexity (1 artificial atom and 1 
resonator). The difference in the distinguishability is even larger if we 
compare to a setup with the same experimental complexity as ours. The setup of 
Ref.~\cite{fan_prb14} with 1~artificial atom, 1~resonator, and 
$\gamma_c/\gamma_{01}=0.1$ reports $\mathcal{F}=84\%$ for $\eta=1$ and 
$\mathcal{F}=75.5\%$ for $\eta=0.5$. With the same $\gamma_c/\gamma_{01}=0.1$, 
our setup achieves $\mathcal{F}=97.6\%$ for $\eta=1$ and $\mathcal{F}=96.9\%$ 
for $\eta=0.5$. This also shows that our setup is much more robust against a 
non-unit homodyne detection efficiency $\eta$ than Ref.~\cite{fan_prb14}.

In conclusion, we propose a nonabsorbing microwave single-photon detector with 
a larger distinguishability than in the similar 
proposals~\cite{sathoyamoorthy_prl14,fan_prb14,royer_prl18}. We accomplish 
this by completely removing the option of the continuous-mode operation 
(without the knowledge of the photon's arrival time) in the proposal of 
Ref.~\cite{fan_prb14} and introducing a new detection sequence. Our proposal 
is different from the previous sequential 
setups~\cite{kono_nphys18,besse_prx18}, as the roles of the artificial atom 
and the resonator are reversed such that the resonator is used as storage, 
instead of the artificial atom. This difference may make our proposal more 
attractive due to often better coherence properties of the resonators compared 
to the artificial atoms.
\begin{acknowledgments}
We thank I. Mahboob and R. Budoyo for useful comments. This work was supported 
by CREST(JPMJCR1774), JST.
\end{acknowledgments}

\bibliography{references}
\end{document}